\documentclass[aps,pre,superscriptaddress]{revtex4}

\usepackage{graphics}
\begin{document}
\def\bskipdm{\mskip -3.6\thickmuskip}
\def\bskiptm{\mskip -3.0\thickmuskip}
\def\bskipsm{\mskip -3.1\thickmuskip}
\def\bskipssm{\mskip -3.1\thickmuskip}
\def\pint{\mathop{\mathchoice{-\bskipdm\int}{-\bskiptm\int}{-\bskipsm\int}{-%
\bskipssm\int}}}
\newcommand{\abs}[1]{\left\vert#1\right\vert}
\newcommand{\imag}{{\rm i}}
\newcommand{\xp}{x^{\prime}}
\newcommand{\Xp}{X^{\prime}}

\begin{center}
\Large\sc
Amplitude equations for systems with long-range interactions
\end{center}

\begin{center}
\large
Klaus Kassner$^{1}$ and Chaouqi  Misbah$^2$
\vskip 0.2 cm
$^1$Institut f\"ur Theoretische Physik,
Otto-von-Guericke-Universit\"at Magdeburg,
Postfach 4120,
D-39016 Magdeburg, Germany\\
$^2$Groupe de Recherche sur les Ph\'enom\`enes hors  de
l'Equilibre, LSP, Universit\'e Joseph Fourier (CNRS),
Grenoble I, B.P. 87, Saint-Martin d'H\`eres, 38402 Cedex, France\\

\vskip 0.2 cm
\vskip 0.2 cm
9 March 2002, optional revision: 26 April 2002

\end{center}

\vskip 1cm

\noindent
We  derive amplitude equations for 
interface dynamics in pattern forming
systems with long-range interactions.
The basic  condition for the applicability of the method developed here
 is that
the bulk equations are linear and
solvable by integral transforms.
We arrive at the interface equation via 
long-wave asymptotics. 
As an example, we treat the Grinfeld instability, and we 
also give a result for the Saffman-Taylor instability.
It turns out that the long-range interaction survives
the long-wave limit and shows up in the final equation
as a nonlocal and nonlinear term, a feature that 
to our knowledge is not shared by any other known
long-wave equation.
The form of this particular equation will then allow us
to draw conclusions regarding the universal dynamics of 
systems in which nonlocal effects persist at the level of 
the amplitude description. 

\vspace{1cm}

\section{Introduction}
A number of  pattern formation problems involving
an interface 
 are computationally difficult despite
the fact that the bulk equations are linear. Examples include
Laplacian dynamics (diffusion-limited aggregation, limit of
vanishing P\'eclet number in dendritic growth), flow problems
in the Stokes approximation, and elastic problems. 
They are 
difficult because of the long-range self-interaction 
of the interface mediated by the external field. With the
availability of modern computer power, these problems can 
usually be solved to good accuracy in two dimensions. But
none of them has been successfully
treated on a large scale in three dimensions
so far. 

Whenever an interface equation can be derived, 
this offers the advantage of better tractability,
both numerically and analytically. Interface equations for
three-dimensional systems are two-dimensional, rendering
much larger systems accessible than bulk simulations 
\cite{kassner98}. For problems containing a length
scale restricting the interaction distance, such as the
diffusion length in solidification problems, interface 
equations are {\em local} in the long-wave limit 
\cite{misbah91,kassner94}, {\em i.e.}, they are partial
differential equations.
The Laplace  or Lam\'e equations,
however, do not have a cutoff distance. 
Hence, it is interesting and important to investigate
how this feature affects the long-wave limit, if any.
There have been interface equations containing nonlocal 
but {\em linear} terms
in the literature before 
\cite{sivashinsky77,hobbs92,marietti01a}. 
As it will turn
out, the nonlocal aspects of the system we consider are
much more ferocious, leading to terms that are both nonlocal 
{\em and} nonlinear.
We are also aware of the case of a previously derived amplitude
equation containing nonlocal nonlinearities \cite{elmer88,christen90}.
However, the equation in question is weakly nonlinear whereas we will
obtain a strongly nonlinear equation.
 
Let us  recall that 
close to the instability
threshold for the merging of  order, systems
exhibiting a type I$_{\rm s}$ instability  in the nomenclature
of Cross and Hohenberg \cite{cross93}, {\em i.e.}, an instability 
at finite wave number, are all expected  
to be described by a universal amplitude 
equation of the Ginzburg-Landau
type. Otherwise, when the critical 
wave number approaches zero (type II$_{\rm s}$)
 the dynamics is described by a long-wave equation such as the
Kuramoto-Sivashinsky (KS) equation.
A prerequisite for these derivations is that nonlocal 
interactions are screened off  
beyond  some  length (small in comparison with $1/q_c$).
When long-range interactions (e.g. in electrostatic, magnetic, 
elastic systems, etc.)
are effectively present, no universal equation has been known as yet
for the latter case. The results of \cite{elmer88,christen90} pertain
to the former case, {\em i.e.}, they constitute a generalization of the 
Ginzburg-Landau description including nonlocal interactions.
We derive here a universal 
form of dynamics  in the long-wavelength limit when 
long-range interactions prevail and obtain a result that is complementary
to the Kuramoto-Sivashinsky description.
 The derivation is 
exemplified on a system undergoing 
a surface instability due to elastic stress. 
Besides the universal feature of the 
resolved question,
we present the practical virtue of this strategy.

Applying our approach to a particular system, 
a uniaxially strained solid undergoing
the Grinfeld instability \cite{grinfeld86,torii92},
a problem which has recently attracted the community of crystal
growth \cite{nozieres93,mueller99,kassner99},
 we obtain an equation that,
when truncated to the leading-order nonlinearities 
should provide an example of the 
universal dynamics. As we shall see, 
this equation contains fewer parameters than
to be expected which is most likely due to the fact that in the case of the Grinfeld 
instability the dynamics is variational. 
Introducing an additional free parameter, we
are able to give the generic form of the universal equation. 
A derivation of this equation
from symmetry considerations will be presented elsewhere \cite{kassner02}.

The paper is organized as follows. In Sec.~\ref{sec:model}, we write down the model
equations for the elastically strained system and introduce the appropriate rescaling
to perform an asymptotic analysis.
Section~\ref{sec:expansion} describes the asymptotic expansion and matching
procedure and presents the final interface equation for two dynamical 
situations corresponding to conservative and to nonconservative dynamics.
In Sec.~\ref{sec:simulation}, we give some simulation results for these 
equations. Section~\ref{sec:universality} contains a discussion of 
universality aspects and gives an example of a simulation of the generic
equation. Finally, we will briefly summarize some conclusions and give an 
outlook in Sec.~\ref{sec:conclusio}.

\section{\label{sec:model} Model equations and nondimensionalization}

The physics of the Grinfeld instability has been discussed in some detail in
\cite{durand98}. We consider the case of a solid in contact with its melt;
let us call the interface position $\zeta(x,t)$ ({\em i.e.}, 
we restrict ourselves to two dimensions). It is related
to the stress  distribution via
  \cite{nozieres93,durand96,durand98,kassner01}: 
\begin{eqnarray}
\frac{\zeta_t(x,t)}{\sqrt{1+\zeta_x^2}} 
&=& -\frac{1}{k \rho_s} \left({1-\nu^2\over 2E}     
   \left[(\sigma_{tt}-\sigma_{nn})^2-\sigma_0^2\right]
   +  \gamma \kappa   +   \Delta\rho g \zeta\right) \>.
 \label{normal_vel}
\end{eqnarray} 
Herein, $1/k$ is a mobility, $\rho_s$ the density of the solid,
$E$ is Young's modulus, $\nu$ the Poisson number,  
$\gamma$ is the surface tension,
$\Delta\rho$ the density contrast between solid and liquid, 
$g$ the gravitational acceleration.
$\kappa=-\zeta_{xx}/\left(1+\zeta_x^2\right)^{3/2}$ 
is the interface curvature, and partial
derivatives of $\zeta$ are denoted 
by subscripts. 
$\sigma_{tt}$ and $\sigma_{nn}$ are 
tangential and normal stresses
 at the interface, {\em i.e.}, $\sigma_{nn}=n_i\sigma_{ij}n_j$,
 $\sigma_{tt}=t_i\sigma_{ij}t_j$, where $n_i$ and $t_i$ are
the components of the normal and the tangent to the interface.
By $\sigma_0$, we denote the
externally imposed uniaxial stress, and the interface position is
measured from the equilibrium position of a planar interface.
(Setting $\kappa=0$ and $\abs{\sigma_{tt}-\sigma_{nn}}=\abs{\sigma_0}$ 
in (\ref{normal_vel}) leads to a steady state $\zeta=0$.) 
Equation (\ref{normal_vel}) 
describes the mass current density which is proportional
to chemical potential changes 
produced by strain, capillarity and gravity.

To solve this equation,
it must be supplemented with the equations describing mechanical equilibrium.
Assuming linear isotropic elasticity,
\begin{equation}
\sigma_{ij}= \frac{E}{1+\nu} \left(u_{ij}
+\frac{\nu}{1-2\nu} \, u_{kk}\, \delta_{ij}\right) \>,
\label{hooke} 
\end{equation}
where $u_{ij}$ is the strain tensor, expressible via the elastic 
displacements $u_i$ according to
$u_{ij} = \frac12\left(\partial_i u_j+\partial_j u_i\right) \>,$ 
the condition for mechanical equilibrium, 
$\partial_j \sigma_{ij}= 0$ (neglecting gravity  in the bulk
as a small cross effect),  
takes the form of the Lam\'e equations:
\begin{equation}
(1-2\nu)\nabla ^2 {\bf u} + {\nabla} ({\nabla}\cdot{\bf u})=0
\; .\label{lame}
\end{equation}
Note that the static version is sufficient since all interface
motions are slow in comparison with sound propagation.
Boundary conditions are
\begin{equation} 
\sigma_{nn}=\sigma_{nt}=0 
\label{siginterf}
\end{equation}
at the interface, periodicity in the $x$ direction with some 
prescribed wavelength $2\pi/q$, and
\begin{equation} 
\sigma_{ij}\to \sigma_0 \,\delta_{ix} \delta_{jx} 
\quad\hbox{for}\quad z\to-\infty\>.
\label{sigminfty}
\end{equation}
Herein, $\sigma_{nt}=
n_i\sigma_{ij}t_j$, is the shear stress
at the interface, which must evidently vanish if the latter is in
contact with a fluid.
Another cross effect has been neglected here, {\em viz.}~that of capillary
overpressure, leading to a curvature dependent jump of $\sigma_{nn}$
across the interface. An overall constant pressure, not influencing
the dynamics of the interface has been subtracted out of the definition of
$\sigma_{nn}$. 

Due to the free boundary character, equations~(\ref{hooke}) through
(\ref{sigminfty}) do not completely specify the problem and  
there is a need for an additional
condition, which is just  equation (\ref{normal_vel}) expressing
the fact that the chemical potential difference between solid
and liquid is proportional to the normal mass current
across the interface. This means that we assume the interface
to be microscopically rough (so that  a linear relation 
holds between the current and its
conjugate variable, the chemical potential difference) and 
the solid to be in contact with its liquid or vapor.

Note however, that for the solution of the elastic problem at one instant
of time with given interface position we do not need (\ref{normal_vel})
(since we have neglected sound propagation effects). This suggests to attack
the problem in two stages: first solve the elastic equations, then move
the interface according to the elastic fields, and repeat the procedure for
the new interface position.

The question is how to obtain a closed interface equation, and to
identify its generic form. In order to proceed, 
we first nondimensionalize
the above  equations 
by introducing two length scales 
$\ell_1 = \gamma E/\sigma_0^2(1-\nu^2)$ 
(the {\em Griffith} length \cite{griffith20},
apart from a factor of $4/\pi$)
and $\ell_2 =  (1-\nu^2)  \sigma_0^2/2\Delta\>\rho\> g E $ (a gravity length)
 as well as a 
time scale $\tau = k\rho_s\ell_1^2/\gamma$  \cite{kassner01}. Then we 
use Hooke's law (\ref{hooke}) to express stresses 
by strains 
and reduce the latter by 
a dimensionless factor $E/(1+\nu)\sigma_0$. 
We find
\begin{equation}
\frac{\zeta_t(x,t)}{\sqrt{1+\zeta_x^2}} =
 -\left\{\frac12 \left[\left(u_{tt}-u_{nn}\right)^2-1\right] 
+\kappa +\alpha\zeta\right\}\>,
\label{nondimevol1}
\end{equation}
where $\alpha=\ell_1/2\ell_2$, {i.e.}~$\alpha\ge0$. 
A linear stability analysis of the full set of equations (\ref{hooke})
 through (\ref{nondimevol1})
 yields the dispersion relation \cite{kassner01}:
\begin{equation}
\omega = 2 \vert q\vert -q^2-\alpha \>,
\label{disprel1}
\end{equation}
where $\omega$ and $q$ are the nondimensional growth rate and wavenumber,
respectively.
The absolute value of $q$ arises, 
because the eigensolutions of the
elastic problem carry a factor $\exp(\pm i q x + \vert q\vert z)$,
approaching zero for $z\to-\infty$. 
A linear instability will arise for $\alpha<1$.

As is obvious from the derivation,  to be able to manifest itself 
the instability requires mass
transport. Besides the case 
described of a 
liquid in contact with its melt, corresponding to  nonconserved dynamics, 
another situation is of particular interest, due to possible applications
in epitaxial growth. This is a solid in vacuum. Material transport
will ordinarily be dominated by surface diffusion then.
Consequently, we have  conserved dynamics, with eq.~(\ref{nondimevol1})
replaced by 
\begin{equation}
\frac{\zeta_t(x,t)}{\sqrt{1+\zeta_x^2}} =
 \frac{1}{\sqrt{1+\zeta_x^2}}\>\partial_x\,
\frac{1}{\sqrt{1+\zeta_x^2}}\>\partial_x\,
\left\{\frac12 \left[\left(u_{tt}-u_{nn}\right)^2-1\right] 
+\kappa +\alpha\zeta\right\}\>,
\label{nondimevol2}
\end{equation}
and the dispersion relation (\ref{disprel1}) is simply multiplied by
a factor $q^2$ on the right-hand side. 
Since the gravity term is normally negligible in
experiments on epitaxial growth (unless a  temperature gradient
mimicks a strong gravitational field \cite{durand96,durand98}),
we may set $\alpha=0$ in this case, obtaining a parameter-free equation.

As it turns out, the asymptotic analysis can be performed together 
for both dynamical situations, since its most extensive part consists
in solving the elastic problem (which is the same for both cases). 
Only afterwards is the solution inserted into the 
equation describing the interface evolution.
At intermediate steps of the calculation, we will therefore write down
equations for the nonconservative case only, 
but we shall give the final interface 
equation for both cases.

Let us now consider a situation, 
where $\ell_1$ is very large, {\em i.e.}, 
we introduce a small
parameter $\epsilon$, setting $\ell_1=O(1/\epsilon)$.
This corresponds to a very small stress, but we can nevertheless reach 
an unstable state by reducing gravity or the density difference, so $\ell_2$
becomes larger than  $\ell_1/2$.

 Then it will be natural to measure
length scales in units of $\epsilon\ell_1$ [which is $O(1)$] 
rather than $\ell_1$ and
time scales in units of $\epsilon^2\tau$. 
Coordinates are transformed according to
$\tilde{x} = x/\epsilon$, $\tilde{z} = z/\epsilon$, 
and $\tilde{t}=t/\epsilon^2$. 
Referring to the new coordinates, 
the dispersion relation transforms into
$\tilde{\omega} = 2 \epsilon \vert \tilde{q}\vert 
                 -\tilde{q}^2-\epsilon^2 \alpha$.
For small $\epsilon$, each term in this relation 
must behave as $\epsilon^2$, hence we have 
for the wavenumber $\tilde{q} \sim \epsilon$. 
A long-wave equation should therefore be
derivable.
To perform the asymptotics, we introduce another set of 
coordinates via $X=\epsilon\tilde{x}$,
$Z=\tilde{z}-\zeta(x,t)$, $T=\epsilon^2\tilde{t}$, 
conveniently mapping the interface 
position to $Z=0$. $X$ and $T$ are slow variables.
We set $U=u_x-(1-\nu) X$, $V=u_z+\nu (Z+\zeta) $,
where $u_x$ and $u_z$ are the displacements,
 which makes the boundary conditions at $-\infty$ 
homogeneous ($U\to 0$, $V\to 0$). 
The transformed  equations (\ref{normal_vel},\ref{lame}) read:
\begin{eqnarray}
&&\epsilon \zeta_T (1+\epsilon^2 \zeta_X^2)^{3/2} = \nonumber\\
&&\mbox{} -\biggl\{ \frac12\biggl[\left(1-\epsilon^2\zeta_X^2\right) 
\left(1+\partial_X U 
-\zeta_X \partial_Z U - \partial_Z V\right) 
+ 2\zeta_X \left(\partial_Z U+\epsilon^2 (\partial_X V
-\zeta_X \partial_Z V)\right)\biggr]^2 \nonumber\\
&&\mbox{}-\frac12 \left(1+\epsilon^2\zeta_X^2\right)^2
-\epsilon \zeta_{XX}(1+\epsilon^2 \zeta_X^2)^{1/2} 
+\epsilon\alpha\zeta(1+\epsilon^2 \zeta_X^2)^2\biggr\} \>, \label{dyneq} 
\end{eqnarray}
\begin{eqnarray}
(1-2\nu) \partial_Z^2 U 
+ \epsilon^2\left[2(1-\nu)\left(\partial_X-\zeta_X\partial_Z\right)^2 U 
+\left(\partial_X-\zeta_X\partial_Z\right)\partial_Z V\right] &=& 0 \>, 
\label{lame1} \\
2(1-\nu) \partial_Z^2 V 
+ \left(\partial_X-\zeta_X \partial_Z \right)\partial_Z U 
+\epsilon^2 (1-2\nu) \left(\partial_X-\zeta_X \partial_Z \right)^2 V &=& 0 \>,
 \label{lame2}   
\end{eqnarray}  
and the boundary conditions at the interface ($Z=0$) become
\begin{eqnarray}
0 &=& (1-\nu) \partial_Z V + \nu \partial_X U 
- (1-\nu) \zeta_X \partial_Z U \nonumber\\
&& \mbox{}
 + \epsilon^2 \left[ (1-2\nu) \left(\zeta_X^2-\zeta_X\partial_X V\right)
+ (1-\nu) \zeta_X^2 \left(\partial_Z V + \partial_X U 
                        - \zeta_X\partial_Z U\right)\right] \>, 
\label{firstbc}\\
0 &=& \left[\partial_Z U 
    + \epsilon^2 \left( \partial_X V- \zeta_X\partial_Z V\right)\right]
 \left(1-\epsilon^2\zeta_X^2\right) 
 + 2 \epsilon^2 \zeta_X \left(\partial_Z V 
  -  \partial_X U +\zeta_X\partial_Z U-1\right)\>. 
\label{secbc}   
\end{eqnarray}
Equations (\ref{dyneq}) through (\ref{secbc}) constitute our starting point
for the asymptotic analysis to be carried out.
 
\section{\label{sec:expansion} Asymptotic analysis}
These equations are singular in $Z$, 
{\em i.e.}, the boundary conditions at infinity,
where the implicit assumption $Z=O(1)$ does not hold anymore,
can be satisfied only with the trivial solution $\zeta_X=0$. 
This difficulty usually does
not appear with equations that have an internal length scale, in contrast to the elastic equations which are devoid of such a scale.
To overcome it, one  introduces an {\em inner} region, 
where $Z=O(1)$, and an {\em outer}
region, where $\eta\equiv\epsilon Z=O(1)$ \cite{hobbs92},
attempting then to match the solutions in their common domain
of validity. 
In the inner region, Eqs.~(\ref{lame1})--(\ref{secbc})
hold, and solving the bulk equations may be reduced 
to ordinary differential equations in $Z$ 
via expansion in powers of $\epsilon$.
In the outer region, derivatives with respect 
to $\eta$ and $X$ are the same order of magnitude.
Therefore, true partial differential equations have to be solved. 
It is here that the linearity
of the Lam\'e equations becomes important, 
because it allows one to solve these equations in
terms of Fourier transforms. Nontrivial aspects of 
the calculation then are largely due to
the asymptotic matching of real space functions with Fourier transforms.

Denoting field variables in the outer region by small letters,
 $u(X,\eta,T)=U(X,Z,T)$, $v(X,\eta,T)=V(X,Z,T)$,
we obtain the {\em outer} equations
\begin{eqnarray}
(1-2\nu) \partial_\eta^2 u 
+2(1-\nu)\left(\partial_X-\epsilon\zeta_X\partial_\eta\right)^2 u 
+\epsilon\left(\partial_X-\epsilon\zeta_X\partial_\eta\right)\partial_\eta v 
&=& 0 \>, 
\label{lame1out} \\
 \left(\partial_X-\epsilon\zeta_X \partial_\eta \right)\partial_\eta u 
+ 2(1-\nu) \epsilon \partial_\eta^2 v
+\epsilon (1-2\nu) \left(\partial_X-\epsilon\zeta_X \partial_\eta \right)^2 v 
&=& 0 \>,
 \label{lame2out}   
\end{eqnarray}  
the {\em inner} ones being given by (\ref{lame1},\ref{lame2}). 
Expanding the interface position and the fields in powers of $\epsilon$,
we find that the expansion is singular in another sense.
If it is assumed that $\zeta=O(1)$,  
nonlinear terms will {\em not} arise in the interface equation,
a fact that can also be inferred from symmetry
considerations and power counting. 
Scalings that lead to a nontrivial
result are $\zeta=O(\epsilon^{-1})$, $u=O(1)$, $v=O(\epsilon^{-1})$. 
We therefore write
\begin{equation}
 \zeta(X,T) =\epsilon^{-1}\zeta_{-1}(X,T)+\zeta_0(X,T)+\epsilon\zeta_1(X,T)+ 
\dots 
\label{expandzeta}
\end{equation}
implying $\partial_{\tilde{x}}\zeta(X,T)=O(1)$.    
Next we set
\begin{eqnarray}
V(Z) &=& \epsilon^{-1}  V_{-1}(Z) + V_0(Z) + \epsilon V_1(Z)+\dots = v(\eta)
     = \epsilon^{-1}  v_{-1}(\eta) + v_0(\eta) + \epsilon v_1(\eta) +\dots,
\nonumber\\
&&\label{expandv}\\
U(Z) &=& U_0(Z) + \epsilon U_1(Z)+\dots = u(\eta)
     =  u_0(\eta) + \epsilon u_1(\eta) +\dots,
\label{expandu}
\end{eqnarray}
where for brevity we have suppressed the $X$ and $T$ dependences.
We then obtain as matching conditions (taking into account the 
$\epsilon$ dependence of $\eta$)
\begin{eqnarray}
 U_{0}(Z) &\to&    u_0(0), \quad (Z\to\infty), \label{matchu0}\\
 V_{-1}(Z) &\to&    v_{-1}(0), \quad (Z\to\infty), \label{matchv-1}\\
 U_1(Z) &\sim& Z u_0^\prime(0) +  u_1(0), \quad (Z\to\infty),
\label{matchu1}\\ 
 V_0(Z) &\sim& Z v_{-1}^\prime(0) +  v_0(0), \quad (Z\to\infty),
 \label{matchv0}
\end{eqnarray}
where the prime denotes a derivative with respect to the argument ($\eta$).
To the two leading orders, the inner equations are solved by
\begin{eqnarray}
U_0(X,Z,T) &=& B_0(X,T) \>, \label{U0sol} \\
V_{-1}(X,Z,T) &=& D_{-1}(X,T) \>, \label{Vm1sol} \\
U_1(X,Z,T) &=& Z A_1(X,T) + B_1(X,T) \>, \label{U1sol} \\
V_0(X,Z,T) &=& Z C_0(X,T) + D_0(X,T) \>. \label{V0sol}
\end{eqnarray}
The outer equations 
need to be solved only at lowest order in $\epsilon$.
For brevity, we rename $\zeta_{-1}(X,T)$ back to  $\zeta(X,T)$.
Confusion should not arise, since we will never need $\zeta_0$
nor $\zeta_1$. It should be kept in mind, however, that the new
$\zeta(X,T)$ is $O(1)$, because it is the prefactor of $\epsilon^{-1}$.
The outer solution then reads
\begin{eqnarray}
u_0(X,\eta,T)  &=& \frac{1}{2\pi} 
       \int_{-\infty}^{\infty} \left[a_0(q,T)+b_0(q,T)(\eta+\zeta(X,T))\right]
                  e^{\imag \, q X+\abs{q}(\eta+\zeta(X,T))} dq\>, 
\label{u0sol} \\[0.2cm]
v_{-1}(X,\eta,T)  &=&  \frac{1}{2\pi} 
            \int_{-\infty}^{\infty} \left\{-\imag\>{\rm sign}(q) 
               \left[a_0(q,T)+b_0(q,T)(\eta+\zeta(X,T))\right] + 
               \imag\>\frac{3-4\nu}{q} b_0(q,T) \right\} \nonumber \\
             &&  \mbox{} \quad e^{\imag \, q X+\abs{q}(\eta+\zeta(X,T))} dq\>.
\label{vm1sol}
\end{eqnarray} 
Using the boundary conditions (\ref{firstbc},\ref{secbc}), 
we can express $A_1$ and $C_0$ by $B_0$ and $D_{-1}$:
\begin{eqnarray}
A_1(X) = \frac{1}{(1-\nu)\left(1+\zeta_X^2\right)^2} &&\left\{\zeta_X\left[
2(1-\nu)+(1-2\nu)\zeta_X^2+((1-\nu)\zeta_X^2+2-\nu)B_0'(X)\right] \right.
\nonumber\\
&&\quad\mbox{}-\left.\left(1-\nu-\nu\zeta_X^2\right)D_{-1}'(X)\right\}\>,
\label{eqa1}\\ 
C_0(X) = \frac{1}{(1-\nu)\left(1+\zeta_X^2\right)^2} &&\left\{\zeta_X^2
+\left(-\nu+(1-\nu)\zeta_X^2\right)
\left(B_0'(X)+\zeta_X(X) D_{-1}'(X)\right)\right\}\>,
\label{eqc0}
\end{eqnarray}
where we have left out the $T$ dependence (which is ``passively'' present
everywhere but important only in the final equation -- the Lam\'e equations
are always solved at a fixed time).
 
The matching conditions at
leading order give four equations expressing $B_0$, $D_{-1}$,
$A_1$, and $C_0$ by Fourier integrals involving $a_0(q,T)$ and $b_0(q,T)$.
For $B_0$ and $D_{-1}$, these integrals are essentially shown in 
Eqs.~(\ref{u0sol},\ref{vm1sol}), we just need to set $\eta=0$ there.
The two remaining equations read:
\begin{eqnarray}
A_1(X)  &=& \frac{1}{2\pi} 
       \int_{-\infty}^{\infty} \Bigl[a_0(q)\abs{q}+b_0(q)(1+\abs{q}\zeta(X))\Bigr]
                  e^{\imag \, q X+\abs{q}\zeta(X)} dq\>, 
\label{foura1} \\[0.2cm]
C_0(X)  &=&  \frac{1}{2\pi} 
            \int_{-\infty}^{\infty} \Bigl\{-\imag q a_0(q)  
               -\imag\>b_0(q)[{\rm sign}(q)+q \zeta(X)] + 
               \imag\>{\rm sign(q)}(3-4\nu) b_0(q) \Bigr\} \nonumber \\
             &&  \mbox{} \quad e^{\imag \, q X+\abs{q}\zeta(X)} dq\>.
\label{fourc0}
\end{eqnarray}

So we have six equations altogether, {\em viz.} Eqs.~(\ref{u0sol})
through (\ref{fourc0}), 
for the six unknown functions 
$B_0$, $D_{-1}$, $A_1$, $C_0$, $a_0$, and $b_0$ \cite{remark01}.
 Their solution determines
the elastic fields, which then can be
inserted into the interface equation (\ref{dyneq}).
However, what we need is an  explicit analytic solution. 
Its details are technical, so we just give some essential
steps.
First, we rewrite the equations in terms of Laplace transforms 
(rather: one-sided Fourier transforms)
\begin{eqnarray} 
\hat{a}_0(x) &\equiv& \frac{1}{2\pi} \int_0^\infty a_0(q) e^{\imag\,qx} dq \>,
\\
\hat{b}_0(x)&\equiv&  \frac{1}{2\pi}\int_0^\infty b_0(q) e^{\imag\,qx} dq \>,
\end{eqnarray}  
for example,
\begin{equation}
B_0(X) = \hat{a}_0(X-\imag\,\zeta(X))+ \zeta(X) \hat{b}_0(X-\imag\,\zeta(X))
+{\rm c.c.} \>.
\end{equation}
 Setting $w=X+\imag\,Z$,
\begin{eqnarray}
2 \hat{a}_0^\prime(w) &\equiv& m(X,Z)+ \imag\,n(X,Z) \>,\\
2 \hat{b}_0(w) &\equiv& o(X,Z)+ \imag\,p(X,Z) \>,
\end{eqnarray}
with $m$, $n$, $o$, $p$ real functions, we can then reduce our problem
 to just two (real) equations involving complex quantities.
These are the Eqs.~(\ref{eqa1},\ref{eqc0})
with all quantities expressed by $\hat{a}_0^\prime(w)$ and by $\hat{b}_0(w)$,
{\em i.e.}, finally by $m$, $n$, $o$, and $p$.

Introducing the abbreviations 
\begin{eqnarray}
\lambda_{{1}}(X)&=&{\frac {\zeta_{{X}}\left (2-2\,\nu+\left (1-2\,\nu
\right ){\zeta_{{X}}}^{2}\right )}{\left (1-\nu\right )\left (1+{\zeta
_{{X}}}^{2}\right )^{2}}}\>, \label{lam1}\\
\lambda_{{2}}(X)&=&{\frac {\zeta_{{X}}\left (2-\nu+\left (1-\nu\right ){
\zeta_{{X}}}^{2}\right )}{\left (1-\nu\right )\left (1+{\zeta_{{X}}}^{
2}\right )^{2}}}\>, \label{lam2}\\
\lambda_{{3}}(X)&=&-{\frac {1-\nu-\nu\,{\zeta_{{X}}}^{2}}{\left (1-\nu
\right )\left (1+{\zeta_{{X}}}^{2}\right )^{2}}}\>,\label{lam3}\\
\lambda_{{4}}(X)&=&{\frac {{\zeta_{{X}}}^{2}}{\left (1-\nu\right )\left 
(1+{\zeta_{{X}}}^{2}\right )^{2}}} \>, \label{lam4}\\
\lambda_{{5}}(X)&=&{\frac {-\nu+\left (1-\nu\right ){\zeta_{{X}}}^{2}}{
\left (1-\nu\right )\left (1+{\zeta_{{X}}}^{2}\right )^{2}}}
\>,\label{lam5}\\  
\lambda_{{6}}(X)&=&{\frac {\zeta_{{X}}\left (-\nu+\left (1-\nu\right ){
\zeta_{{X}}}^{2}\right )}{\left (1-\nu\right )\left (1+{\zeta_{{X}}}^{
2}\right )^{2}}}\>,\label{lam6}
\end{eqnarray}
these two equations become
\begin{eqnarray}
&& {\mskip-5\thickmuskip}
\Bigl[1-\lambda_2(X)\zeta_X+(3-4\nu)\lambda_3(X)\Bigr] \, o(X,-\zeta)
+\Bigl[-\lambda_2(X)+\lambda_3(X)\zeta_X\Bigr] \, m(X,-\zeta)
\nonumber\\
&&=\>\mbox{}\lambda_1(X) 
+ \Bigl(\lambda_2(X)-\lambda_3(X)\zeta_X\Bigr)\, \zeta  \, o_X(X,-\zeta)
\nonumber\\ 
&&\quad\mbox{}\Bigl(-1+\lambda_2(X)\zeta_X+\lambda_3(X)\Bigr)\left(n(X,-\zeta)
+\zeta \,p_X(X,-\zeta)\right) -2(1-2\nu)\lambda_3(X)\,\zeta_X \,p(X,-\zeta) \>,
\nonumber\\
&&\label{longeq1}\\
&&{\mskip-5\thickmuskip}
\Bigl[-1-\lambda_5(X)+\lambda_6(X)\zeta_X\Bigr] \, m(X,-\zeta)
+ \Bigl[-\lambda_5(X)\zeta_X+(3-4\nu)\lambda_6(X)\Bigr] \, o(X,-\zeta)
\nonumber\\
&&=\>\mbox{}\lambda_4(X) 
- \Bigl(-1-\lambda_5(X)+\lambda_6(X)\zeta_X\Bigr)\,\zeta \, o_X(X,-\zeta)
+2(1-2\nu)\Bigl(1-\lambda_6(X)\zeta_X\Bigr) \, p(X,-\zeta) 
\nonumber\\
&&\quad\mbox{}+
\Bigl(\lambda_5(X)\zeta_X+\lambda_6(X)\Bigr)\left(n(X,-\zeta)
+\, \zeta  \, p_X(X,-\zeta)\right)\>,
\label{longeq2}
\end{eqnarray}
where $p_X$ and $o_X$ denote the  derivatives of  $p(X,-\zeta)$,
$o(X,-\zeta)$ with respect to $X$ at fixed $\zeta$. That is, we first take
the derivative and only then allow $\zeta=\zeta(X)$.

Since the four functions involved are real and imaginary parts of 
 of analytic functions, only two of them are independent. Indeed, we   
have
\begin{eqnarray}
n(X,-\zeta) &=& -\frac{1}{\pi} \int_{-\infty}^{\infty} 
                \frac{(\Xp-X)\> m(\Xp,0)}{(\Xp-X)^2+\zeta^2}\, d\Xp\>, 
\label{relrealimag1}\\ 
p(X,-\zeta) &=&  -\frac{1}{\pi} \int_{-\infty}^{\infty} 
                 \frac{(\Xp-X)\> o(\Xp,0)}{(\Xp-X)^2+\zeta^2}\, d\Xp\>. 
\label{relrealimag2}
\end{eqnarray}
Equations (\ref{longeq1}) and (\ref{longeq2}), being linear, can be
solved formally for $o(X,-\zeta)$ and $m(X,-\zeta)$. 
This leads to
\begin{eqnarray}
o(X,-\zeta) &=& r_1(X)\left[n(X,-\zeta)
+\zeta p_X(X,-\zeta)\right] + r_2(X) p(X,-\zeta) +r_3(X)\>,
\label{oformal}\\
m(X,-\zeta) &=& r_4(X)\left[n(X,-\zeta)
+\zeta  p_X(X,-\zeta)\right] + r_5(X) p(X,-\zeta) +r_6(X)\zeta o_X(X,-\zeta)
\nonumber\\
&&\mbox{} + r_7(X) \>,
\label{mformal}
\end{eqnarray}  
where the functions $r_1$ through $r_7$ are given by
\begin{eqnarray}
r_{{1}}(X)&=&\frac{1}{1-2\,\nu}\>,\\
r_{{2}}(X)&=&{\frac {2\,\zeta_{{X}}}{(1-2\,\nu)\,(1+\zeta_{{X}}^{2})}}\>,\\
r_{{3}}(X)&=& {\frac {-\zeta_{{X}}}{(1-2\,\nu)\,(1+\zeta_{{X}}^{2})}}\>,\\
r_{{4}}(X)&=&0\>,\\
r_{{5}}(X)&=&2\,{\frac {-1+\nu\,(1+{\zeta_{{X}}}^{2})}{1+{\zeta_{{X}}}^{
2}}}\>,\\
r_{{6}}(X)&=&-1\>,\\
r_{{7}}(X)&=&-{\frac {{\zeta_{{X}}}^{2}}{1+{\zeta_{{X}}}^{2}}}\>.
\end{eqnarray}

The computations leading from (\ref{u0sol})
through (\ref{fourc0}) to (\ref{oformal},\ref{mformal}) are
straightforward but
 heavy. Some intermediate steps ({\em e.g.}, the simplification
of the $r_{{i}}(X)$) as well as a verification of the whole calculation
at the end have been done using computer algebra (MAPLE). 
Due to the chosen scalings (\ref{expandzeta}) -- (\ref{expandu}),
Eqs.~(\ref{oformal},\ref{mformal}) constitute an exact
reformulation of the outer elastic problem, which is linear in $u$ and $v$,
so higher order equations have the same form as the leading order
in $\epsilon$. To obtain analytic results, we have to employ some
approximation. We assume that $\zeta(X)$, the expansion coefficient
of $\epsilon^{-1}$, is itself a numerically small quantity. 
Then expressions such 
as (\ref{relrealimag1},\ref{relrealimag2}) become {\em Hilbert transforms}. 
Denoting the Hilbert transform of $f(X)$ by  
\begin{equation}
{\cal H}\Bigl[f(X)\Bigr] \equiv \frac{1}{\pi} 
\pint_{-\infty}^{\phantom{-}\infty} \frac{f(\Xp)}{\Xp-X}\, d\Xp\>,
\end{equation}
where the integral is to
be taken as a principal value, as indicated by the  bar,
we find, as the limit $\zeta\ll 1$ of (\ref{oformal},\ref{mformal})
\begin{eqnarray}
o(X,0) &=& -r_1(X)\,{\cal H}\Bigl[m(X,0)\Bigr]  - r_2(X)\,{\cal H}\Bigl[o(X,0)\Bigr] +r_3(X)\>,
\label{ohilbert}\\
m(X,0) &=&  - r_5(X)\,{\cal H}\Bigl[o(X,0)\Bigr]  + r_7(X) \>.
\label{mhilbert}
\end{eqnarray}  
This set of linear equations can be solved iteratively. 
Assuming, and this is the second approximation, that $\zeta_X(X)$ is also
numerically small, we may truncate the interation after the first
step, taking accurately into account only terms up to order $\zeta_X^2(X)$.  
Using ${\cal H}\Bigl[{\cal H}[f(X)]\Bigr]=-f(X)$, we can solve the truncated
iteration to obtain explicit expressions for $o(X,0)$, $m(X,0)$
that read:
\begin{eqnarray}
o(X,0) &=& -(1-2\nu)\,r_3(X) + {\cal H}\Bigl[r_7(X)\Bigr] \>,
\label{osol}\\ 
m(X,0) &=& -(1-2\nu)\,r_7(X) -2(1-\nu)(1-2\nu) {\cal H}\Bigl[r_3(X)\Bigr]\>.
\label{msol}
\end{eqnarray}   
Substituting $o$, $m$, $n=-{\cal H}[m]$, and $p=-{\cal H}[o]$
back into the equations for $A_1$, $C_0$, $B_0$, and $D_{-1}$,
inserting these into (\ref{U0sol}) -- (\ref{V0sol}) and then back into
(\ref{dyneq}), we finally obtain 
 the sought-for interface equation, in which we rename $X$ into $x$
 and $T$ into $t$
(they are the same variables anyway):
\begin{eqnarray}
\frac{\zeta_t(x,t)}{(1+\zeta_x^2)^{1/2}} &=&
    \frac12 \left\{ 1 - \frac{1}{(1+\zeta_x^2)^2} \left[ (1-\zeta_x^2) 
    \left(1+\frac2\pi \pint_{-\infty}^{\phantom{-}\infty}
          \frac{\zeta_x(\xp)}{(\xp-x)(1+\zeta_x(\xp)^2)} d\xp \right) 
          + \frac{4 \zeta_x^2}{1+\zeta_x^2}\right]^2 \right\}\nonumber\\
&& \mbox{} +\frac{\zeta_{xx}}{(1+\zeta_x^2)^{3/2}}-\alpha \zeta \>.
\label{interfeq}
\end{eqnarray}
The arguments of $\zeta$ ($x$ or $\xp$ and $t$) have
been suppressed wherever possible without ambiguity.

 The nonlocal term drives
the instability.
It is interesting
to see whether (\ref{interfeq})  will reproduce a
 {\em coarsening dynamics} of the type discussed in \cite{kassner01}.
For in that case whether
a groove grows or shrinks  
does not depend on local quantities (such as the
curvature) alone. Hence, the scenario described
cannot be expected from a merely local equation.
 
This is the melt-crystallization case \cite{kassner01}. It is easily checked 
that linearization of this equation produces the 
linear dispersion relation (\ref{disprel1}).
The corresponding equation with
dynamics controlled by surface diffusion is obtained  simply by preceding
the right-hand side with the operator 
$-\left[(1+\zeta_x^2)^{-1/2} \, \partial_x\right]^2$:
\begin{eqnarray}
\zeta_t(x,t) &=&
\partial_x\,\frac{1}{\sqrt{1+\zeta_x^2}}\>\partial_x\, \left(
    \frac12 \left\{ \frac{1}{(1+\zeta_x^2)^2} \left[ (1-\zeta_x^2) 
    \left(1+\frac2\pi \pint_{-\infty}^{\phantom{-}\infty}
          \frac{\zeta_x(\xp)}{(\xp-x)(1+\zeta_x(\xp)^2)} d\xp \right)
\right.\right.\right.
\nonumber\\ 
 && \mbox{} + \left.\left.\left. 
\frac{4 \zeta_x^2}{1+\zeta_x^2}\right]^2 \right\}
 -\frac{\zeta_{xx}}{(1+\zeta_x^2)^{3/2}}+\alpha \zeta
\vphantom{ \left[ (1-\zeta_x^2) 
    \left(1+\frac2\pi \pint_{-\infty}^{\phantom{-}\infty}
          \frac{\zeta_x(\xp)}{(\xp-x)(1+\zeta_x(\xp)^2)} d\xp \right) 
          + \frac{4 \zeta_x^2}{1+\zeta_x^2}\right]}
\right) \>.
\label{interfeqcons}
\end{eqnarray}
For the system described by this equation, coarsening dynamics does not 
yet seem   to have been studied in the literature, 
therefore it will be briefly discussed in the next  section. 
 
\section{\label{sec:simulation} Some simulation results}

\begin{figure}[h]
\resizebox{0.7\textwidth}{!}{\includegraphics{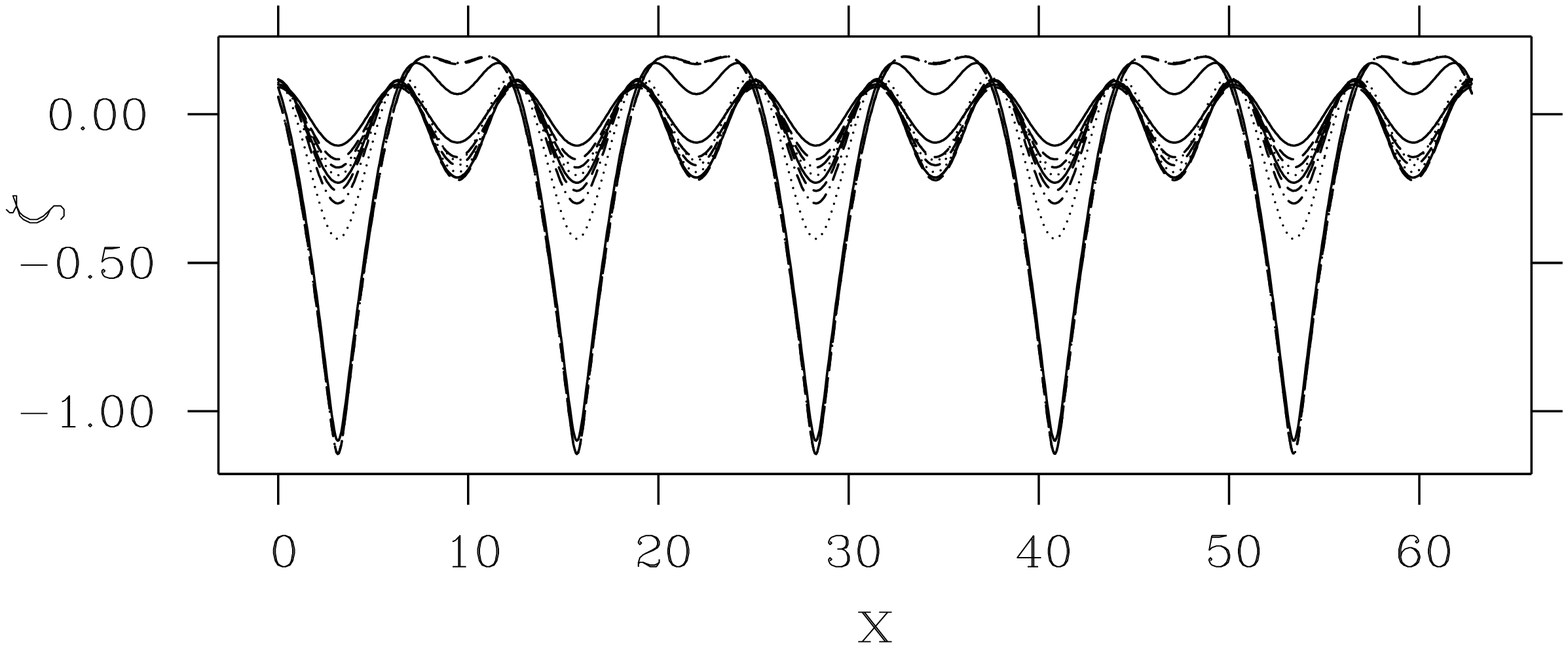}}
\caption{Period doubling for the interface profile. 
The initial interface is a cosine with wavelength $2\pi$ and amplitude 0.1,
with every odd minimum made deeper by 5\%.
$\alpha=0.97$. Time interval between curves: $\Delta t =10.0$.}
\label{perioddoubncns}
\end{figure}

\begin{figure}[h]
\resizebox{0.7\textwidth}{!}{\includegraphics{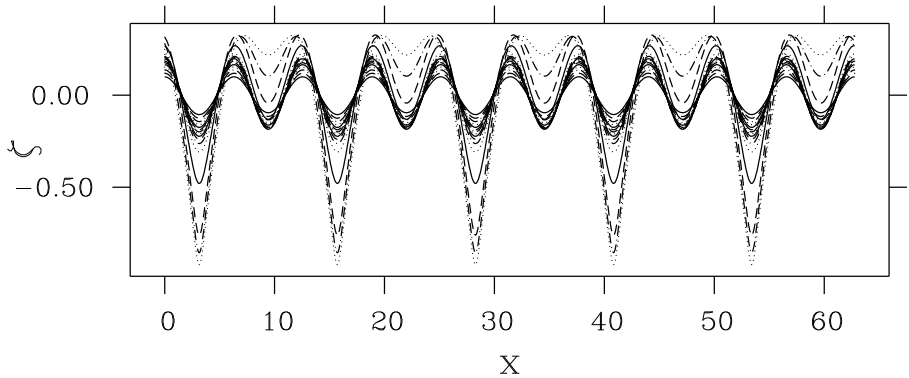}}
\caption{ 
The initial interface is a cosine with wavelength $2\pi$ and amplitude 0.1,
with every odd minimum made deeper by 5\%.
$\alpha=0.97$. Time interval between curves: $\Delta t =10.0$.}
\label{perioddoubcons}
\end{figure}

Both equations (\ref{interfeq}) and (\ref{interfeqcons})
have been simulated numerically. Figure \ref{perioddoubncns}
 gives an example
for period-doubling dynamics found in the nonconservative dynamics
when an initially periodic sinusoidal
structure is submitted to a small perturbation of twice 
its basic periodicity.
A similar result is obtained for conservative dynamics and the same
parameter $\alpha$, see Fig.~\ref{perioddoubcons}. In both cases, $\alpha$
was close to 1, hence the instability was weak. For smaller values of
$\alpha$, we found stronger differences between the behavior of the
two systems. The main statement that can be made, however, is that in
both of them similar coarsening scenarios are operative, since coarsening
allows a reduction of the overall elastic energy. Because of
the restriction that the average interface height has to remain constant
in the conservative case, the latter has coarsening proceed more slowly,
as can be verified by comparison of Figs.~\ref{perioddoubncns} and 
\ref{perioddoubcons}. 

\begin{figure}[h]
\resizebox{0.7\textwidth}{!}{\includegraphics{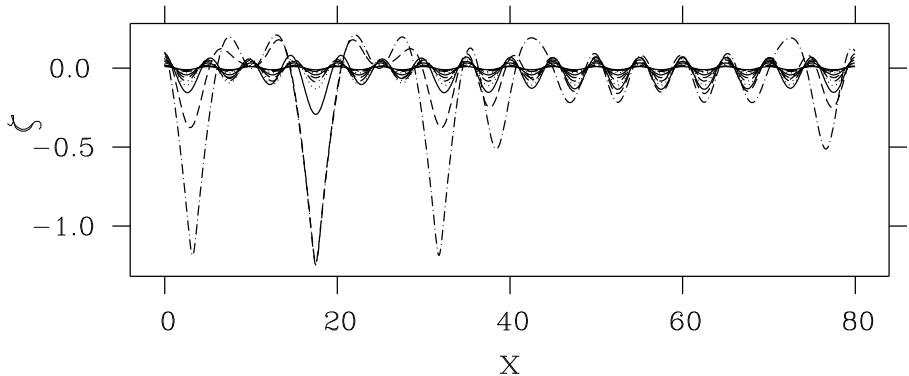}}
\caption{Coarsening of the interface profile. The fourth groove
of the initial interface has been made deeper by 1\%.
$\alpha=0.90$. Initial
wavelength $2\pi$. Time interval between curves: $\Delta t =10.0$.}
\label{coarsening}
\end{figure}

In Fig.~\ref{coarsening}, we give the dynamics obtained from a 
{\em local} perturbation
of a periodic interface. The process of coarsening is 
more complex than the simple period-doubling scenario suggested in 
\cite{kassner01}. Frustration effects come into play when grooves
compete via the nonlocal interaction, so it may happen that not the
next-nearest neighbor of the leading groove is a winner of the competition
but rather the third neighbor.   

An extensive overview of the properties of 
Eqs.~(\ref{interfeq},\ref{interfeqcons}) is not intended here. It will
be the subject of a different publication \cite{kassner02b}.
Suffice it to say  that we find these amplitude equations to
reproduce several
 qualitative features of the dynamics of the system faithfully, 
including {\em nonlocally induced coarsening}. 
Differences with the full dynamics
\cite{kassner01} appear at large amplitudes. For vanishing gravity
({\em i.e.}, $\alpha=0$), the amplitude  is not predicted to 
saturate in the full model but
does so in the amplitude equation. 
This is not so surprising given  that the latter
was derived assuming small amplitudes (more precisely
a small prefactor of the $O(\epsilon^{-1})$ term in the amplitude 
expansion).
In principle, systematic improvement 
is possible by continuing the iteration scheme truncated after the
first step and including higher powers of $\zeta$ and  $\zeta_x$.
In practice, this improvement results in additional technical difficulties.

It is certainly not possible to claim genericity of
Eqs.~(\ref{interfeq},\ref{interfeqcons}). Their derivation
accounts for only the lowest-order ({\em i.e.}, quadratic) nonlinearities
exactly, while resumming some nonlinearities  to infinite order, 
which are presumably
specific for the elastic system.
However, if there is a universal amplitude equation for the
systems considered, then it must contain the terms that our derivation
produces exactly, and it may be sufficient to consider just these terms.

\section{\label{sec:universality} Generic equations}

Before looking at the generic equation(s) derivable from our 
particular case, let us briefly recapitulate the general situation 
concerning  amplitude equations of universal validity.

There is an over-abundance of nonequilibrium systems 
which spontaneously build up 
an organized  pattern from an initially structureless 
state when they are taken sufficiently
far from equilibrium \cite{cross93}. 
Typical examples are present in  hydrodynamical systems 
({\em e.g.}, Rayleigh-B\'enard
convection), chemical reactions ({\em e.g.}, Turing systems), 
crystal growth (the Mullins-Sekerka
instability), and so on. 
Despite the fact that the underlying physical and chemical
mechanisms are diverse in these systems, sufficiently 
close to the instability
threshold they all fall within the same universality class.  
Let $q_c$ denote the wavenumber
of the emerging ordered pattern, and  $h(x,t)$ a field describing
the pattern (which can stand for a component
of the velocity field in hydrodynamics, or an interface 
position in crystal growth, etc...) along
the $x$ direction.
Close to the threshold it is known \cite{cross93} 
that $h\sim A(x,t)e^{iq_cx}$, where $e^{iq_cx}$ describes
the (rapid) periodic variation of the field due to the birth of order, 
while $A$ is a slowly
varying (slow with respect to $e^{iq_cx}$) amplitude. 
$A$ obeys the following equation 
\begin{equation}
A_t=A+A_{xx}-  |A|^2A \>, 
\label{amplitude}
\end{equation}
where  coefficients can always be set
to unity by an appropriate rescaling. 
This equation is  universal in the sense that its form depends only
on translational and rotational symmetries. 
It is usually referred to as the {\it amplitude equation} or the 
Ginzburg-Landau equation \cite{cross93}. 
The amplitude equation describes the nature of the bifurcation
from the structureless state as well as instabilities of 
the ordered pattern with respect
to wavelength modulations (the Eckhaus instability \cite{cross93}).

A description in terms of a slowly varying amplitude makes a sense only 
if $A$ varies in a slow fashion in comparison with the pattern 
wavelength $1/q_c$. This requirement fails
if $q_c\rightarrow 0$. In that case a separation between a 
fast and slow variation
is illegitimate. One has thus to resort to a (singular) expansion \`a la 
Sivashinsky \cite{sivashinsky77}.
Because $q_c$ is small, it provides an appropriate 
parameter of expansion. This is also
known as the {\it long-wavelength limit}. 
In that limit the field $h$ obeys generically
the Kuramoto-Sivashinsky equation (or the damped form 
with $\lambda$ the damping rate), which is written in the canonical form,
\begin{equation}
h_t=-\lambda h -h_{xx}-h_{xxxx}+ h_x^2 \>.
\label{KS}
\end{equation}
Again this equation is universal \cite{misbah94a,remark} 
in the limit $q_c\rightarrow 0$.
Because of the smallness of $q_c$ (long-wavelength) 
any nonlocal effect present in the original
constitutive equations disappears, and one recovers
again a local equation. 
This is true only because a length scale 
is present and serves as a cut-off. It is akin to the Debye length
which makes the nonlocal response function of the Coulomb gas local.
There is, however, a myriad of situations where 
long-range interactions play a decisive
role at all scales, and where it is hardly
 believable that the above equation (or a similar equation)
should arise even in the long-wavelength limit. 
Typical situations with long range interactions
are electrostatic and magnetic systems, 
problems with elastic fields, and so on. These systems
are devoid of an intrinsic length-scale and 
it is highly desirable to derive the appropriate
generic form of the corresponding evolution equation 
in the long-wavelength limit. A derivation on the basis of
a gradient expansion of an appropriately chosen general nonlinear
operator in the spirit of \cite{misbah94a} will be 
given  elsewhere \cite{kassner02}.

Here we argue that we can essentially read off the form of that equation from
our result (\ref{interfeq}). To this end, we simply expand to second order
in $\zeta_x$ and drop all higher-order terms. As noted before, the kept
terms are exact to that order. The   
 ``generic'' evolution equation 
with long range interaction obtained this way reads
\begin{equation}
\zeta_t = -\alpha \zeta + \zeta_{xx} 
- \frac2\pi \pint_{-\infty}^{\phantom{-}\infty}
                  \frac{\zeta_x(\xp)}{\xp-x} d\xp -2\zeta_x^2
                  -{1\over 2} \left( 
                  \frac2\pi \pint_{-\infty}^{\phantom{-}\infty
                  }
                                    \frac{\zeta_x(\xp)}{\xp-x} d\xp
                                    \right)^2 \>.
\label{nonlocal}
\end{equation}
Comparing this equation with the KS one (\ref{KS}),
we see that it is not exactly a generalization. Since the nonlocal
term drives the instability, the diffusive term need not be negative
and hence there is no need for a fourth-order term. 
Rather, Eq.~(\ref{nonlocal}) is {\em complementary} to the  Kuramoto-Sivashinsky
result. Of course, the analog of Eq.~(\ref{nonlocal}) with a negative
sign of $\zeta_{xx}$ may also appear and then the dispersion relation
would need a different ultraviolet stabilization. If this term were
fourth order in $q$, we would obtain a linear term $\propto \zeta_{xxxx}$
and an equation that is a common generalization of the KS equation and
Eq.~(\ref{nonlocal}). 

Closer inspection shows that this equation can be only a particular case of 
the generic equation. For in general the amplitude equation will
have coefficients in front of each of its six terms. Four of these
can be made equal to one via division by a common prefactor and
rescaling of space, time and the amplitude
$\zeta$ itself. So our final equation should contain two independent
nondimensional parameters, but it has only one, $\alpha$.
We suspect that this is due to a hidden symmetry, and a natural candidate
for this symmetry is the variational nature of the basic evolution
dynamics (\ref{normal_vel},\ref{lame}). In the fully generic equation,
the additional parameter would be the ratio of the coefficients of the
two nonlinear terms.
It is therefore easy to conjecture the form of this equation:
\begin{equation}
\zeta_t = -\alpha \zeta + \zeta_{xx} 
- \frac2\pi \pint_{-\infty}^{\phantom{-}\infty}
                  \frac{\zeta_x(\xp)}{\xp-x} d\xp -2\zeta_x^2
                  -{\beta\over 2} \left( 
                  \frac2\pi \pint_{-\infty}^{\phantom{-}\infty
                  }
                                    \frac{\zeta_x(\xp)}{\xp-x} d\xp
                                    \right)^2 \>,
\label{nonlocalgen}
\end{equation}    
where we have introduced a second parameter $\beta$.
The corresponding universal equation for conservative dynamics
would be obtained by preceding the complete right-hand side of
this equation by the operator $-\partial_x^2$.

\begin{figure}[htb]
\resizebox{0.8\textwidth}{!}{\includegraphics{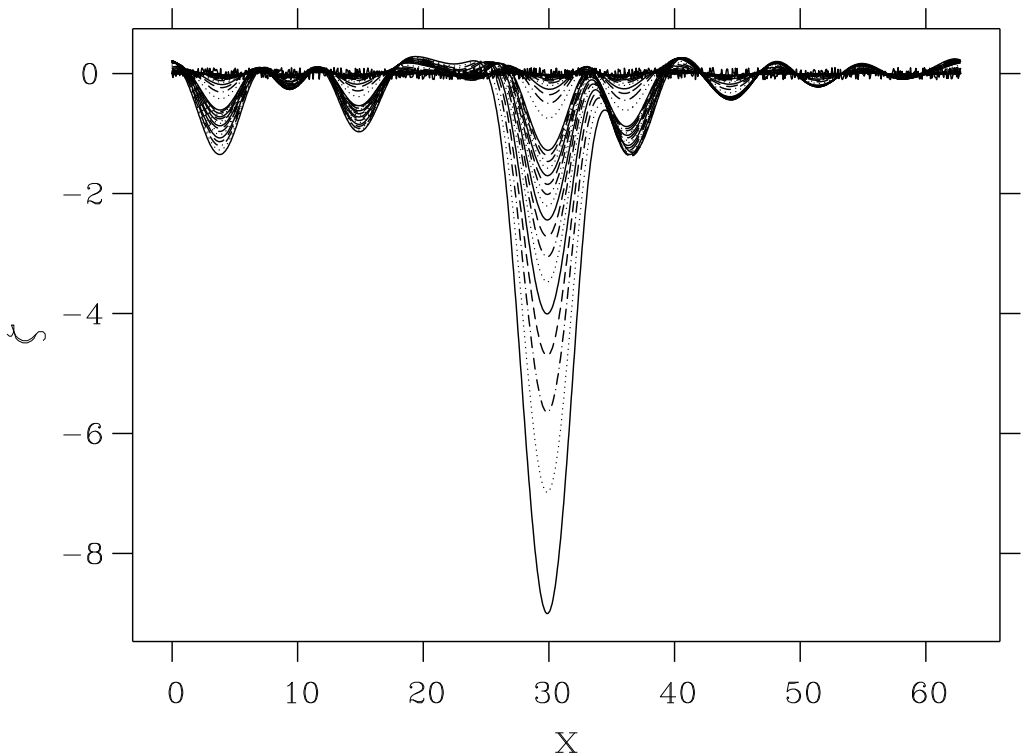}}
\caption{Evolution according to the generic equation 
(\protect\ref{nonlocalgen}). $\alpha=0.8$,
$\beta=1.0$. The initial interface is a random structure (visible
near $z=0$).
 Time interval between curves: $\Delta t =1.0$ up to $t=20.0$,
thereafter $\Delta t =0.1$, and the evolution is shown up to $t=21.6$.
Note the spectacular acceleration of the ``winning groove''.}
\label{geneqsim}
\end{figure}

We have simulated Eq.~(\ref{nonlocal}) and find that it still describes
coarsening, this time without saturation. An example of a simulation is
displayed in Fig.~\ref{geneqsim}. We see a single groove or ``finger''
survive after a long time.

On physical grounds, an non-saturating amplitude is to be expected
in some situations involving the Grinfeld instability \cite{kassner01}.
Nevertheless, even in such a case
one should  {\em a priori} not believe that
a long-wave equation predicting an ever-increasing
amplitude correctly describes the long-time asymptotics, 
because the interface eventually leaves
the domain of validity of the long-wave assumption.
This happens when the amplitude becomes larger than the wavelength.
Therefore, an equation such as (\ref{nonlocalgen}) should be 
expected to describe  some intermediate-time asymptotics at best.

Things are different for other universal equations such as 
the Ginzburg-Landau equation (\ref{amplitude}) and the 
(damped) Kuramoto-Sivashinsky
equation (\ref{KS}), which usually do work in the long-time limit.
But this is due to the fact that in these cases the final 
(average) amplitude
does saturate and is predicted to be small. However, if the cubic
term in  (\ref{amplitude}) is positive, which necessitates a 
fifth-order describtion to obtain saturation of the amplitude,
the final asymptotics cannot ordinarily be trusted either, because
then the amplitude may not be small contrary to the original 
assumptions.

On the other hand, while any amplitude description leading to 
non-saturating amplitudes may seem self-defeating in this sense,
there appear to be systems, where the domain of validity of
the amplitude equation extends beyond that of the assumptions
made in its derivation. Recently, a singular long-wave equation was
derived for step-flow growth without desorption \cite{pierrelouis98}.
It predicts an amplitude growing as the square root
of time, {\em i.e.},  without bounds. Currently available evidence
on the basis of numerical solutions of the exact model equations
seems to corroborate that prediction. Hence, it is imperative
to compare the long-time limit of the amplitude description to
whatever information is available about the long-time behavior
of the full system, in order to assess its validity.

Equation (\ref{nonlocalgen}) should be thought of as
being the first in a hierarchy of generic equations describing
the initial and intermediate stages of evolution
 of a number of different systems. It is relevant
to systems the linear dispersion relation of which is quadratic.
For other systems, such as the Saffman-Taylor system, where the
dispersion relation is cubic, we will have a different linear
term in the equation (the Laplacian of a Hilbert transform),
and it is an interesting question whether the appearing nonlinearities
will remain the same.   
Following the  same strategy for the Saffman-Taylor problem, we arrive
at the following equation
\begin{equation}
\zeta_t =   
-\frac{V}{\pi}\pint_{-\infty}^{\phantom{-}\infty} 
\frac{\zeta_x(\xp)}{\xp-x} d\xp
-\frac{\gamma B}{\pi} \pint_{-\infty}^{\phantom{-}\infty} 
\frac{\zeta_{xxx}(\xp)}{\xp-x} d\xp 
+ V \zeta_x^2 + \gamma B \zeta_x \zeta_{xxx} \>,
\end{equation}
where $V$ is the flat-interface velocity, $\gamma$ the surface tension,
and $B=b^2/12\mu$ with $b$ the thickness of the 
Hele Shaw cell and $\mu$ the fluid viscosity.
Of course the linear terms are always quite trivial, but what is
noteworthy is that $\beta=0$, {\em i.e.,}
the nonlinear terms are local, like in the KS equation. 
(Under the usual assumptions for long-wave equations, the second
local nonlinearity is small in comparison with the KS one.)
We call this limit the Sivashinsky limit where nonlocality is present
in the linear terms only, as was derived originally by Sivashinsky for flame
propagation \cite{sivashinsky77}. This contrasts with our 
equation (\ref{nonlocalgen}) which involves in the long-wavelength limit
a {\it nonlinear and nonlocal} term. Therefore, this equation
should introduce a new universality class with long range interactions.

\section{\label{sec:conclusio}conclusions}

To summarize, our method of derivation is applicable whenever the 
general solution of the bulk equations
can be found by a transformation method.
Usually this means they must be {\em linear} (with constant coefficients). 
A second condition, requiring that the dispersion relation can
be transformed to long-wave form, should be always satisfiable.
The generalization to three dimensions is straightforward, in principle.
Our approach opens up the road to a realm
of amplitude equations constituting a hitherto unknown type,
the investigation of which seems worthwhile
both from the mathematical and physical points of view; 
 a new line of research should be stimulated. 

The fact that the inner expansion 
produces only polynomials in $Z$ (and hence the outer solution
is identical to the {\em uniform} approximation \cite{Bender78})
strongly suggests that the derivation may even be achieved
without the detour via a long-wave approach, 
using a  regular perturbation expansion of
the outer equations. This would seem to indicate
that we are not really dealing with an equation that is valid only
in the long-wave limit
but  rather with an ordinary (but {\em strongly nonlinear})
amplitude equation valid at all wavelengths
for small enough amplitudes, thus {\em a posteriori} justifying 
the  formal device of rescaling to large wavelengths.
On the other hand, the first of our approximations corresponds to
setting $\abs{q}\zeta$ equal to zero in the
exponents of (\ref{u0sol}) and (\ref{vm1sol}), 
which, after taking into account all rescalings, means
that the amplitude is small
compared with the wavelength.

Finally, the amplitude equation is much faster (and easier) 
to handle numerically than previous schemes. 
Like the KS equation, for example,
its leading behavior
described by Eq.~(\ref{nonlocalgen}) should be generic
for systems undergoing long-wavelength instabilities with long-range
interactions.

{\bf Acknowledgment} We are grateful to Peter Kohlert for help with the 
computer algebra.

\newcommand{\phre}{Phys. Rev. E}


\begin{thebibliography}{10}
\bibitem{remark01}
Actually 
$a_0$ and $b_0$ are complex while the other quantities are real, but
they have to satisfy $a_0^\ast(q) =a_0(-q)$ and $b_0^\ast(q) =b_0(-q)$,
where the star denotes complex conjugation, to make
the Fourier transforms in Eqs.~(\protect\ref{u0sol},\protect\ref{vm1sol}) 
and (\protect\ref{foura1},\protect\ref{fourc0}) real.
\bibitem{remark} This is true generically under the condition that 
no constraint, such as conservation laws for $h$, is imposed and that
the system is nonvariational. Still,
the equation has a partial differential form.
\end{thebibliography}
\end{document}